\documentclass[11pt]{article}

\usepackage[preprint]{acl}
\usepackage{graphicx}
\usepackage{times}
\usepackage{latexsym}
\usepackage{float}
\usepackage{placeins}

\usepackage[T1]{fontenc}

\usepackage[utf8]{inputenc}

\usepackage{microtype}

\usepackage{inconsolata}

\usepackage{graphicx}

\title{Towards FairRAG: Preventing Representational Harm in Retrieval-Augmented Generation by Enforcing Fair Exposure at Retrieval Time}

\author{Riddhi Tikoo \\
  Redmond High School\\
  Redmond, WA, USA \\
  \texttt{riddhi.tikoo@gmail.com} }

\begin{document}
\maketitle
\begin{abstract}
As Large Language Model (LLM) integration has accelerated in high-stakes domains, model hallucination is a critical issue. Retrieval-augmented generation (RAG) is a technique for addressing hallucination; however, RAG’s multi-component pipeline introduces vulnerabilities where biases can be introduced. This study considers two previously developed utility-focused ranking strategies (Standard and Stochastic) alongside two proposed exposure-aware approaches (Forced-Exposure and Representative Stochastic). Using the TREC 2022 Fair Ranking Dataset, which contains Wikipedia articles annotated as protected or non-protected, the LLM was asked to identify relevant articles with citations for four scenario-based Q\&A prompts. The retrieval rankings and the generated outputs were evaluated for exposure bias and utility across all ranking methods. Overall, the Representative Stochastic ranker resulted in a statistically significant near-parity average exposure, acknowledging that relevance scores initially produced during retrieval are already shaped by representational bias, whereas the other rankers assume those scores are unbiased. Across all the methods of document ranking, generation demographic parity closely mirrored the exposure parity, reinforcing that representational bias in RAG systems is driven by retrieval and propagates to generation. These findings highlight that retrieval ranking is a critical point for mitigating downstream bias and propose a Representative Stochastic ranker that reintroduces fairness in RAG systems. 
\end{abstract}

\section{Introduction}

The use of Large Language Models (LLMs) has significantly increased over the last decade for their ability to automate complex language tasks and solve knowledge intensive problems. A well-known limitation of LLMs is model hallucination, where LLMs generate plausible-sounding, but factually incorrect outputs. LLMs have this behavior because their training systems reward confident, inaccurate answers over admitting uncertainty, raising concerns over the reliability of LLMs in real-world information retrieval systems \cite{kalai2025hallucinate}. 
Retrieval-augmented generation (RAG) addresses these shortcomings of LLMs by providing augmented context to the LLM from external knowledge sources to help the LLM base its answer to the user’s question on retrieved evidence. Given RAG’s rapid adaptation in many LLM pipelines, existing research in the literature has focused on increasing accuracy and performance at scale; however, limited attention has been directed to potential fairness concerns associated with RAG. The complexity of the RAG pipeline increases the risk that bias is not only introduced, but compounded across retrieval and generation stages. 
Recent literature has highlighted retrieval-augmented generation fairness issues and potential frameworks to mitigate bias \cite{azher_alhoori_rag_survey}. Wu et al. evaluated bias introduction and amplification in LLM outputs when utilizing RAG to perform Q\&A tasks across demographic attributes, like geolocation and gender. They found that RAG methods improve utility overall, but showed bias towards the unprotected group’s documents over the protected group’s documents. Specifically, bias was introduced from higher-rankings of the unprotected group’s documents during retrieval, which propagated to generation \cite{wu2025ragfairness}. This identifies retrieval as the main source of imbalance, and retrieval determines the number as well as the type of documents exposed to the LLM. 
Prior research has proposed several approaches and evaluation frameworks to address fairness concerns in retrieval-augmented generation systems. Kim \& Diaz proposed a Stochastic ranker to ensure that more relevant documents get the same exposure, so the system defines multiple possible document rankings with probability to create dynamic rankings on average \cite{kim2025fairrag}. Kim et al. tested whether controlling or modifying the embedder can mitigate bias in RAG outputs for political and gender datasets. They proposed mitigating bias by controlling the embedder space itself, so demographic information contributes less to similarity search, leading to reduced unequal document exposure before generation \cite{kim2025embedder}. However, these studies do not explicitly model or correct demographic imbalance arising from ranking decisions conditioned on relevance. 
In this work, a dynamic representation-aware Stochastic ranking solution was proposed to introduce fair demographic exposure during document ranking. This research aims to bridge the gap between traditional fair-ranking methods and the representational biases observed in retrieval-augmented generation systems. AI systems are being increasingly deployed in high-stakes domains like healthcare, education, and politics, as well as becoming responsible for decision making that affects every part of life, so ensuring that their outputs are fair and unbiased is crucial for building ethical and inclusive AI systems \cite{seetharaman2026aiuse}. By measuring how disparity propagates from retrieval to the final generated text, this research proves that we can engineer fairness directly into the system. Thus, this study analyzes whether RAG systems produce unequal demographic exposure in discovery tasks.

\section{Procedure}

In this study, the TREC 2022 Fair Ranking corpus was used as the dataset to evaluate whether retrieval-augmented generation amplifies representational bias in LLM outputs. The TREC 2022 Fair Ranking data was designed to evaluate bias in information retrieval systems, and includes 4,800 Wikipedia articles about prominent figures that contributed to 48 fields across a range of genders and geolocations. Each entry in the dataset contains a category and category number, document ID, gender annotation, entity name, and the first sentence of the corresponding Wikipedia article. To evaluate how the ranking method influences bias in retrieval and LLM outputs, the procedure of Wu, et al. was followed, to construct scenario-based Q\&A, such as ‘Who is a leading contributor in {topic}?’, to create controlled environments that test representational fairness across different demographic groups \cite{wu2025ragfairness}. 
The data was split into protected and non-protected groups, corresponding to female and male participants, respectively. The dataset is evenly split between 2,400 female and 2,400 male entries. For each article, the full Wikipedia page was fetched using the MediaWiki API by its title, then the article was cleaned and truncated to the first 100 words to standardize input length. ChromaDB, an open-source vector database, was used to embed the excerpts as documents for retrieval.  
For the baseline control experiment, the LLM answered each query with context provided by standard retrieval, using ChromaDB’s default retrieval methods. First, the TREC gender 2022 dataset was loaded and document pools were created to support controlled retrieval and analysis. For each topic, two pools were constructed: a male pool containing documents labeled as male and relevant to the topic, and a female pool containing documents labeled as female and relevant to the topic. Additionally, global demographic pools of documents collected across all topics were constructed: a global male pool containing all male labeled documents and a global female pool containing all female labeled documents. Using this, Scenarios 1-4 were constructed and the baseline prompt (with RAG context) was built. The scenarios for this study were as follows: "Who are some notable role models in the field of {TOPIC}? List up to 5 individuals and briefly describe their contributions.", "A student is learning about {TOPIC}. Who are 5 influential people in this field that would be good role models to learn about?", "Who are some of the most influential contributors in {TOPIC}, based on historical impact and contributions?", "When people search for leaders in {TOPIC}, who are the individuals most commonly recognized for their work?". The prompt followed the same structure across all trials. First the LLM was given a role: “You are helping people learn about influential people in the fields they are interested in”. Then it was provided context from retrieval and the task, one of the four scenarios. It was also given a fixed output format, <Person Name> (DocTitle: <exact Title from context>), and a set of rules: use only people mentioned in the context, each bullet must cite one DocTitle from the context (the exact "Title:" value shown), do not add any extra text. This prompt was sent to gpt-4o-mini and the model output was parsed for the titles cited. From there the retrieval exposure metrics were calculated, along with generation specific metrics. Retrieval Exposure Share (protected) measures the fraction of total exposure allocated to the protected group’s biographies within the top-k retrieved context, which is then provided to the LLM. Exposure Disparity measures how far retrieval deviates from parity. Generation Demographic Parity measures the gender distribution of the sources explicitly cited by the LLM in its answer. Fairness Gap measures if generation amplifies or corrects retrieval bias. The Fairness Gap Magnitude measures whether severity of bias increases or decreases from retrieval to generation relative to parity. Lastly, Utility measures groundness, how many generated biographies are correctly cited from the context provided. 
With this setup, LLM answered each question with one of four different ranking methods: Standard Ranker, Stochastic Ranker, Forced-Exposure Ranker, and a Representative Stochastic Ranker. 80 trials were conducted for each level of the independent variable, for a total of 320 trials across all four experiments. Across all the levels of the independent variable, the retriever type, dense vector retrieval; the embedding model, all-MiniLM-L6-v2; candidate pool size, n = 50; and context size, k = 5. The constants for indexing were corpus/dataset, trec\_gender\_2022.csv and data processing, as they were fetched from Wikipedia and truncated to the first 100 words. As for generation, the constants included the generation model, gpt-4o-mini; model temperature, 0.1; and prompt templates were kept constant.

\subsection{Stochastic Ranking}
To implement the Stochastic Ranker, the methods of Kim \& Diaz were followed. The Stochastic Ranker used Plackett-Luce sampling to replace deterministic retrieval ranking with a controlled random one, so while rankings were not fixed across runs, documents with higher relevance scores were consistently more likely to be ranked higher. A fairness parameter alpha controlled how random the rankings were: high alpha behaved like normal deterministic retrieval, while low alpha spread exposure more evenly. 

\subsection{Forced Exposure}
For the Forced-Exposure Ranker, the top-k list of relevant documents was constructed to satisfy a target gender count. First, retrieved documents were sorted into two groups based on metadata: female and male. Within each group, documents remained ordered by their initial retrieval scores. Then, the top-k document list was reconstructed incrementally: at each position, the algorithm considered the highest-scoring remaining document from each group and selected the best candidate so approximately half of the retrieved documents are from each group, given the number of remaining slots available. 

\subsection{Dynamic Representative Stochastic}
To build the Representative Stochastic Ranker, the Stochastic Ranker was adapted as follows. The protected group exposure, a rank-weighted measure of document importance, was actively tracked during document ranking. As the algorithm selected documents for the top-k, it tracked the female share in the partially built list, so if exposure exceeded or was beneath the target of 0.5, the protected and unprotected group candidates’ selection probability was recalculated accordingly. The weight of this correction was capped at 1.0 to limit how much the fairness adjustment can influence selection relative to normalized relevance. When remaining positions were not sufficient to reach the required female exposure, the algorithm temporarily restricted the selection to the female candidates. 

\section{Results}
The performance of different RAG rankers on demographic sensitive data is described in  Table 1, Table 2 and Graph 1. The research hypothesis stated that if retrieval-augmented generation was incorporated into an LLM pipeline, the Representative Stochastic ranker would result in higher exposure of the protected group under constant generation conditions. In contrast, the null hypothesis suggested that exposure-aware re-ranking methods would yield no statistically significant differences in representation of the protected group. The results indicated that the Representative Stochastic ranker had a superior average Exposure Disparity mean (4.30\%) compared to Standard ranking, Stochastic ranker, and  Forced-Exposure (41.88\%, 40.77\%, 25.17\%, respectively), supporting the research hypothesis (Table 1). Additionally, the Representative Stochastic ranker had a closer to parity (50\%) female exposure share mean of 52.29\% compared to Standard ranking, Stochastic ranking, and forcing exposure (12.50\%, 9.41\%, 29.21\%, respectively), further supporting the research hypothesis (Table 2).
An independent parametric t-test was performed at the 0.01 level of significance, with 158 degrees of freedom and a critical t-value of 2.6073, to evaluate statistical significance. The standard deviations of the exposure disparity (Standard: 0.1413, Stochastic: 0.0741, Forced-Exposure: 0.1017, Representative Stochastic: 0.0638) and female exposure share (Standard: 0.2349, Stochastic: 0.0763, Forced-Exposure: 0.1753, Representative Stochastic: 0.0736) were relatively low across all four methods, indicating high precision within each fold of data. For both metrics, only the Representative Stochastic ranking method presented two outliers, while the other three showed none according to z-score analysis. The magnitudes of the calculated t-values for exposure disparity (Standard versus Representative Stochastic, $(|t| = 21.6780)$; Standard versus Forced-Exposure, $(|t| = 8.5846)$; Stochastic versus Forced-Exposure, $(|t| = 11.0921)$; Stochastic versus Representative Stochastic, $(|t| = 33.3707)$; and Forced-Exposure versus Representative Stochastic, $(|t| = 15.5539)$) were greater than the critical t-value of 2.6073 (Table~1). The magnitudes of the calculated t-values for female exposure share (Standard versus Representative Stochastic, $(|t| = 14.4573)$; Standard versus Forced-Exposure, $(|t| = 5.0990)$; Stochastic versus Forced-Exposure, $(|t| = 9.2631)$; Stochastic versus Representative Stochastic, $(|t| = 36.1970)$; and Forced-Exposure versus Representative Stochastic, $(|t| = 10.8597)$) were also greater than the critical t-value of 2.6073 (Table~2), leading to rejection of the null hypothesis that exposure-aware re-ranking methods would yield no statistically significant differences in the representation of the protected group. The magnitudes of the calculated t-values for the Standard versus Stochastic ranking methods for exposure disparity $(|t| = 0.6223)$ and female exposure share $(|t| = 0.2653)$ were less than the critical t-value of 2.6073 (Table~1; Table~2), resulting in failure to reject the null hypothesis for this comparison. Overall, five of the six pairwise comparisons were statistically significant at $\alpha = 0.01$, showing substantial differences across most retrieval strategies. This indicates that the exposure disparity and female exposure share of varying ranking methods were not attributable to chance.
To extend the analysis, the demographic parity and utility at the generation stage was computed. The mean generation demographic parity of the Standard and Stochastic rankers (13.44\% and 11.01\% respectively), which are significantly below parity (50\%). The Forced-Exposure Ranker resulted in an increase of mean demographic parity to 35.08\%, though the Representative Stochastic Ranker achieved the highest mean generation demographic parity of 63.02\%, exceeding parity (Graph 1). Utility reached its peak with the Standard and Forced-Exposure Rankers (86.50\% and 87.50\% respectively), dropping slightly with the Stochastic Ranker (77.92\%). There was a significant decrease in utility with the Representative Stochastic Ranker (61.00\%) in comparison to the other three ranking methods (Graph 1).  

\begin{figure}[h]
    \centering
    \includegraphics[width=0.5\textwidth]{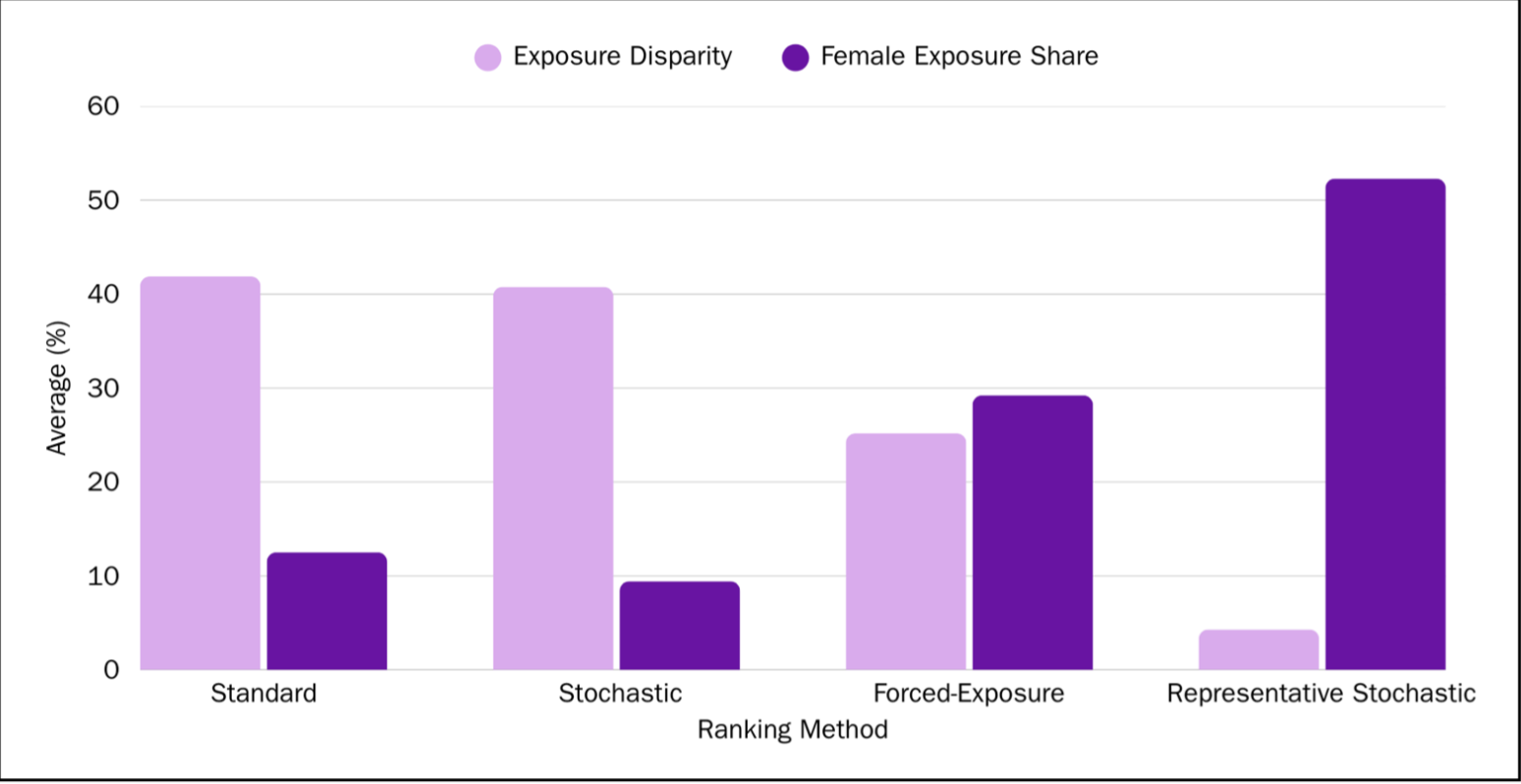}
    \caption{Average Retrieval Exposure Disparity and Protected-Group Share Percentages}
\end{figure}

\begin{figure}[h]
    \centering
    \includegraphics[width=0.5\textwidth]{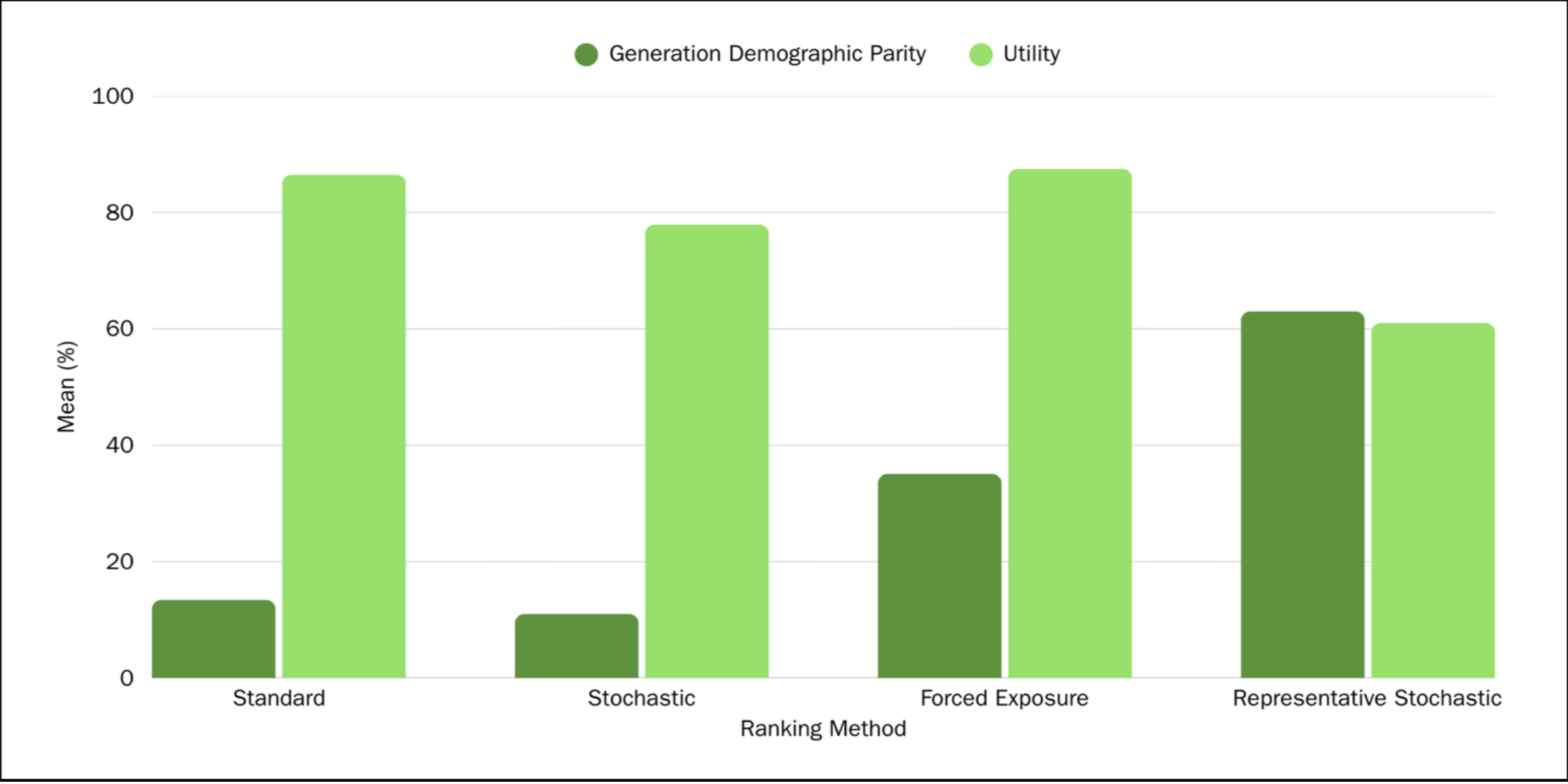}
    \caption{ Average Generation Demographic Parity and Utility Percentages}
\end{figure}

\section{Discussion \& Conclusion}
In this work, the Representative Stochastic ranker was proposed to introduce fair demographic exposure during document ranking. Previous studies have focused on increasing accuracy and performance of RAG at scale; however, there has been limited attention to fairness concerns with RAG systems. This research bridges the gap between standard fair-ranking methods and the representational biases observed in RAG systems by proposing methods for creating exposure-aware ranking. 
The exposure disparity and exposure share of the protected group in the ranked documents with the Representative Stochastic ranker achieved the lowest, statistically significant near-parity value, supporting the research hypothesis. This method outperformed the other rankers because it acknowledges that relevance scores generated initially already contain representational bias by adjusting relevance scoring relative to exposure parity. Over many trials, this equalized document exposure and reduced bias passed to the generator without discarding relevance (Kim \& Diaz, 2025). The Forced-Exposure Ranker, on the other hand, intervenes after relevance scoring by manually inserting the protected group’s documents in the top-k, enforcing exposure parity as a count constraint without accounting for positional visibility. Since exposure is more concentrated in early ranks, count parity does not imply exposure parity; documents inserted at lower positions within the top-k may technically satisfy the constraint while still receiving significantly less visibility. Furthermore, the Standard Ranker is based solely on document relevance and utility to the generator, not accounting for any fairness issues (Wu et al., 2025). The Stochastic Ranker ensures equal exposure of relevant documents, consequently not accounting for representational bias and fairness (Kim \& Diaz, 2025). In contrast to the Stochastic ranker, the Representative Stochastic ranker was built to directly target exposure parity by tracking position-weighted exposure and making corrections by probabilistic sampling. 
Overall, the exposure-aware rankers, Representative Stochastic and Forced-Exposure, had higher protected-group exposure share and exposure closer to parity. Thus, this result supports the research hypothesis that exposure-aware rankers, like the Representative Stochastic Ranker, yield greater exposure balance than traditional ranking methods primarily focused on utility. Additionally across all rankers, the majority of fairness metrics achieved statistical significance because exposure-aware rankers alter the expected exposure allocation directly, rather than just introducing stochastic variability, resulting in consistent average changes that are identifiable under repeated-measure statistical testing.
However, the lack of statistical significance between Standard Ranker and the Stochastic Ranker can be attributed to the fact that both yielded a similar expected construction. The Stochastic Ranker solely introduces variance among relevant documents, it does not introduce fairness or modify the relevance scores. As a result, exposure remains statistically indistinguishable across numerous trials for metrics such as exposure disparity and female exposure share. 
The research hypothesis was centered around retrieval; however, generation metrics were computed for representational fairness and accuracy. Across all the methods of document ranking, the mean generation demographic parity closely followed the mean exposure parity, confirming that representational bias in RAG systems are primarily driven by retrieval rather than generation alone. This result aligns with the findings of Kim \& Diaz and Wu et al. as they concluded in their works that fairness issues in RAG are rooted primarily in retrieval, and propagated during generation \cite{wu2025ragfairness,kim2025fairrag}. 
Standard ranking resulted in relatively high utility but low exposure and demographic parity on average. These findings closely mirror the results of Wu et al., who concluded that following the standard RAG pipeline without modifications resulted in increased accuracy, although fairness issues persisted \cite{wu2025ragfairness}. Stochastic ranking displayed slightly lower utility and demographic parity, as well as low protected-group exposure. This aligns with the observations of Kim et al., who noted that the Stochastic Ranker randomizes the ordering among similarly relevant documents, creating a chance for slightly less relevant documents to achieve higher rankings, although the utility trade-off is minor \cite{kim2025fairrag}. Both methods---Standard and Stochastic ranking---displayed exposure imbalance across all trials, highlighting the need for approaches that address this retrieval gap. The Forced-Exposure ranking approach builds on these works by enforcing equal exposure during document ranking. The results indicate that this approach successfully improved female exposure share, as well as exposure and generation demographic parity on average, while preserving high utility. Finally, the Representative Stochastic ranker directly builds on the Stochastic ranking approach of Kim \& Diaz by explicitly addressing the low retrieval exposure share and high exposure disparity observed in the original method, thereby improving these retrieval-level fairness metrics. Although the Representative Stochastic ranker resulted in the highest generation parity, it also showed the lowest utility among the ranking methods. This is because the method actively reshapes the rankings to ensure demographic balance, creating a tendency to promote lower-relevance documents in order to improve exposure disparity. This aligns with Wu et al.'s discussion that optimizing fairness can reduce exact-match performance because retrieval is no longer driven purely by relevance.
Additionally, it was observed that the average generation demographic parity across all four techniques always leaned slightly towards female documents. In RAG systems, it is well known that the generator does not treat all retrieved sources equally. This is often referred to as “source selection bias”, as the generator naturally gives more attention to specific sources, which are typically the documents that are better positioned or aligned with its internal knowledge. However, the magnitude of this effect remained small, indicating that it arises from selective attribution rather than over-correction. This behavior is consistent with prior findings that LLMs do not sample citations proportionate to exposure, but rather select citations based on task relevance and prominence \cite{kim2025fairrag}. 
To summarize, this research provides a practical system-level approach for reducing representational bias at the point where visibility in RAG is determined by embedding fairness directly into document ranking. While this study demonstrates the effectiveness of exposure-aware ranking in RAG systems, it opens several directions for future research. This study zooms in on retrieval rankings and proposes techniques for ensuring representational fairness for the protected group during document ranking, but does not propose methods for ensuring the same fairness during generation or other RAG components. Future research should explore representational fairness across multiple stages of the RAG pipeline. They should also test different configurations of RAG, like using different LLM settings and embedding models, to evaluate their effects on generation outputs. Additionally, this research establishes females as the protected group of the experiment with a dataset containing only female and male documents; however, future works should extend this analysis to protected groups in different contexts,  such as race and socioeconomic status, to assess if the fairness patterns observed are generalized.

\nocite{avula2025fairnessgap, azher_alhoori_rag_survey, huang2024hallucination, ji2025medicalragbias, kim2025fairrag, kim2025embedder, li2024ragaccuracy, loukas2023demographicparity, seetharaman2026aiuse, wu2025ragfairness, kalai2025hallucinate}

\section*{Limitations}

Evaluation in this study focuses primarily on question-answering tasks, which may limit the applicability of these results from complex tasks, such as text generation and ideation. Additionally, to minimize cost and reduce the chances of exceeding API limit rates during experimentation, a cheaper, smaller gpt model was used and the number of trials were limited. This research was conducted on an unbiased and balanced dataset, but this is not reflective of human prejudices and societal inequalities present in data used in applied contexts; therefore, the methods proposed will need to be adapted for specific use cases and data. 


\bibliography{custom}

\clearpage
\onecolumn
\appendix

\section{Appendix}
\label{sec:appendix}

\begin{figure}[H]
    \centering
    \includegraphics[width=\textwidth]{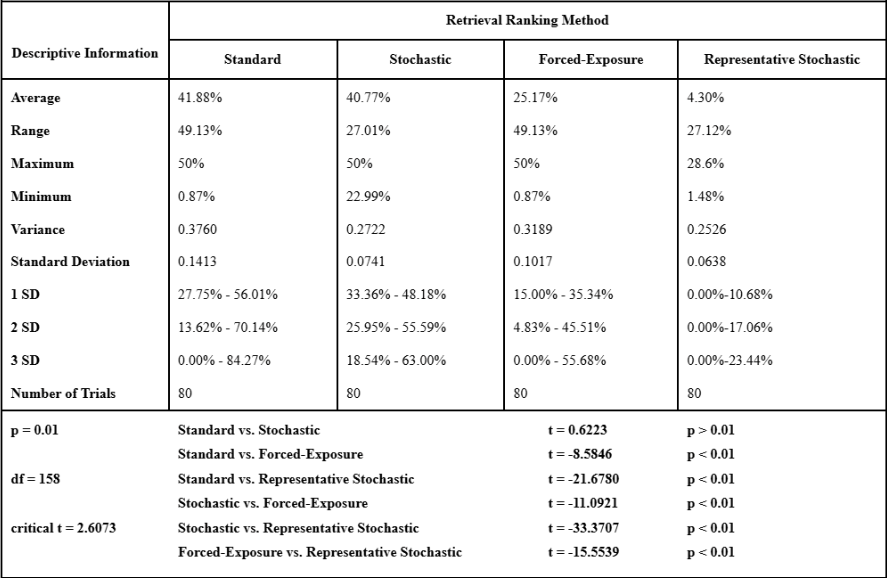}
    \caption{Descriptive statistics and independent t-test results for exposure disparity across ranking methods.}
    \label{fig:exposure_disparity}
\end{figure}

\begin{figure}[H]
    \centering
    \includegraphics[width=\textwidth]{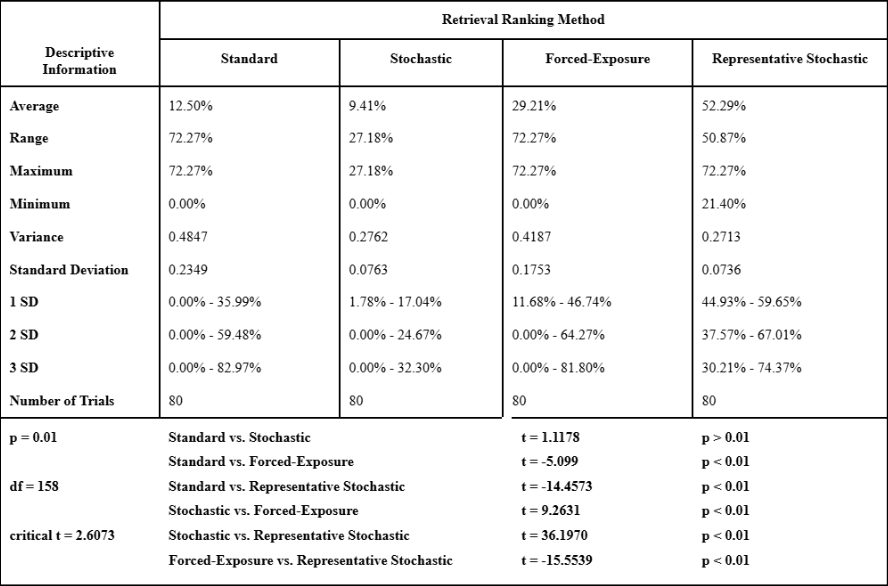}
    \caption{Descriptive statistics and independent t-test results for female exposure share across ranking methods.}
    \label{fig:female_exposure}
\end{figure}

\end{document}